# Two-dimensional heterogeneous photonic bandedge laser

Soon-Hong Kwon,[a], Se-Heon Kim, Sun-Kyung Kim, and Yong-Hee Lee

*Department of physics, Korea Advanced Institute of Science and Technology,*

*Taejon 305-701, Korea*

Sung-Bock Kim

*Telecommunication Basic Research Laboratory, Electronics and Telecommunications*

*Research Institute, Taejon 305-600, Korea*

*Abstract–We proposed and realized a two-dimensional (2D) photonic bandedge laser surrounded by the photonic bandgap. The heterogeneous photonic crystal structure consists of two triangular lattices of the same lattice constant with different air hole radii. The photonic crystal laser was realized by room-temperature optical pumping of air-bridge slabs of InGaAsP quantum wells emitting at 1.55 $\mu$m. The lasing mode was identified from its spectral positions and polarization directions. A low threshold incident pump power of 0.24 mW was achieved. The measured characteristics of the photonic crystal lasers closely agree with the results of real space and Fourier space calculations based on the finite-difference time-domain method.*

[a] Electronic mail: ksunong@kaist.ac.kr

Since Eli Yablonovitch and Sajeev John firstly demonstrated the concept of photonic bandgap materials in 1987, [1, 2] they have attracted much attention as new building blocks in photonics because of its ability to control photons in the scale of wavelength. Specially, various photonic crystal (PC) lasers have been studied by many groups worldwide owing to demand of submicron sized coherent sources for photonic integrated circuits. In order to fulfill the conditions for efficient small sources, i.e., high quality factor, low threshold, and high coupling ratio with waveguides, various types of cavity structures like single defect,[3] ring type,[4] stick type,[5] slab edge,[6] and coupled cavities,[7] have been proposed and studied. In the most of PC lasers, PC functions as a mirror. On the other hands, several groups have taken notice of possibility that the flat dispersion curve near photonic bandedge in the band structure can be used for lasing operation.[8-14] The group velocity near the bandedge is zero because of coupling between the Bloch waves, and then photons in the PC can interact with an active material for a long time, which result in lasing operation. Recently, photonic bandedge lasers have been realized in the organic material,[13] semiconductor quantum wells.[8-11] We and Monat et al. have shown the lasing operation of the photonic bandedge mode in the sample with small active area ~ (10×10 μm) using 2D slab structure.[9-11] However, in the real sample, the group velocity of the bandedge mode becomes nonzero due to finite size of the PC pattern[14] so that in-plane loss, i.e., the propagation of the mode out of the pattern through the slab, is unavoidable. On the other hands, the bandedge mode below light line is vertically well-guided inside the slab. Therefore, if the in-plane loss can be reduced, the bandedge mode would be a mode with very low loss.

In this letter, we propose a new cavity structure based on the bandedge mode. In order to prevent in-plane loss, photonic bandgap region can be introduced as a mirror outside the PC region, in which photonic bandedge mode operates. If the PC pattern of outside region is different from the bandedge mode operating region so that the frequency of the bandedge mode lies within the photonic bandgap of surrounding structure, then the in-plane loss of the bandedge mode would be inhibited. PC pattern can be modified by changing of several parameters such as lattice constant, air hole radius, and lattice geometry. We changed the air hole radii of the surrounding structure in our PC pattern. Here, we investigate the characteristics of the photonic bandedge laser surrounded by the photonic bandgap wall in free-standing slab triangular lattice structures at room temperature. Because the photonic bandgap assists the lasing of the bandedge mode, we call this type of laser the photonic bandgap-bandedge (PBG-BE) laser. We identified the lasing mode from the measurement of the spectral position and polarization characteristics, and compare them with theoretical calculation based on three-dimensional (3D) finite-difference time-domain (FDTD) method.

Six pairs of InGaAsP quantum wells with an emission peak at 1.55 μm are embedded in the center of the slab. We constructed triangular lattice structures by electron beam lithography and $Cl_2$-assisted Ar-beam etching techniques. The InP sacrificial layer below the slab was removed using a diluted HCl : $H_2O$ (4:1) solution through the etched air holes to form free-standing slabs. The thickness of the slab is 282.5 nm, which supports only the fundamental transverse electric (TE)-like mode. Figure 1(a) shows a scanning electron microscope image of PBG-BE laser sample. Air holes in the central region of the pattern have relatively larger radii (0.35$a$) than that of the surrounding air holes (0.30$a$). This cavity structure can be understood that the hexagon with five larger holes along the ΓK directions is placed inside the defect cavity, in which five holes are missing along the ΓK directions. Since the normalized frequency of the band structure increases with respect to the air hole radius, the frequency of the K point bandedge mode operating in the central region is placed within the photonic bandgap of the outside region, as shown in Fig. 1(b). In the experiments, the air hole radii of the central region were designed to be in the range of 0.26$a$ ~ 0.38$a$ and outside holes were kept to be smaller radii as 0.05$a$ than that of the central region. In order to confine the bandedge mode horizontally, 11 smaller air holes are placed along the ΓK directions from the boundary of the bandedge mode region. The lattice constants range from 325 nm to 414 nm with about 20 nm steps to match the gain region of the InGaAsP quantum wells into the first K point bandedge. The fabricated samples with various parameters were pulse-pumped using a 980 nm laser diode at room temperature. The pulse width was 7.5 ns and the duty cycle was 0.1 % to avoid thermal heating effect. The pumping laser was focused with microscope objective lens and the emission of light was collected with the same lens positioned vertically to the sample surface. The pumping spot diameter is about 3.5 μm.

A typical PBG-BE mode pattern in the lasing sample was captured by the charge-coupled device (CCD), as shown in Fig. 2(a). The bright spot was observed within the central region with larger air holes, which means that the lasing mode is localized by the photonic bandgap. In addition, the state of polarization was measured to confirm that the lasing mode is originated from the first K point bandedge mode. Because one-dimensional (1D) distributed feedback (DFB) bandedge modes such as the first M, second M points, consist of a pair of anti-parallel wave vectors, the polarizations become linear, as Ref. 11 shows. On the other hand, the oscillations of the first K point bandedge mode result from the coupling of several nonparallel wave vectors.[13] Therefore, the mode is expected to be unpolarized. In fact, the measured lasing mode shows no definite polarization direction, as shown in Fig. 2(b). In order to further identify the lasing mode, the measured normalized frequencies (plotted as dots) for each lattice constant are compared with the calculated frequencies (plotted as lines) as a function of air hole radius in the central region, as shown in Fig. 2(c). The size of bandedge operating region is finite so that the normalized frequency of the lasing mode differs slightly from it of the first K point bandedge due to uncertainty of the wavevectors. Thus, the plane-wave-expansion (PWE) method can not be used to calculate the exact frequency of the PBG-BE mode. Therefore, in order to find the exact spectral position, the calculated frequencies in the Fig. 2(c) are obtained through the 3D FDTD method. The measured spectral positions agree well with the theoretical predictions over broad range of air hole radii for each lattice constant. The difference between simulations and experiments is less than 1 % of the normalized frequency, which corresponds to the error ~ $0.01a$ of air hole radius measurement.

To analyze the PBG-BE mode theoretically, we compare the PBG-BE mode and the bandedge mode operating at the sample without surrounding air holes through the electric field intensity profiles obtained by the 3D FDTD method. In this simulation, air hole radius of central region, and that of outside region were $0.35a$ and $0.30a$, respectively. In the top-view and the side-view of the intensity profile of Fig. 3(a) and (b), the electric fields of the PBG-BE mode are well localized at the central region. On the other hands, in the case of the bandedge mode operating at the undressed finite sample where there are no surrounding air holes, significant fields are propagating through the slab out of the PC pattern, as shown in Fig. 3(c). In the point of loss mechanisms, the role of the surrounding air holes becomes clearer. In the PC pattern including 21 air holes along the ΓK direction, which is large enough for lasing, the vertical and horizontal quality factors of the undressed bandedge mode are 37000 and 1200, respectively, i.e. the in-plane loss is dominant. Here, the definition of the vertical and horizontal quality factor is followed by the Ref. 14. Although the quality factor of the bandedge mode increases as the number of air holes, [14] the PBG-BE mode operating in the PC pattern with only five air holes along the ΓK direction has relatively large quality factor ~50000. This high quality factor results from the fact that the photonic bandgap prohibits the in-plane loss of the bandedge mode. Thus, we can conclude that the surrounding PC pattern with smaller air holes functions as a mirror for the bandedge mode. On the other hands, since this hetero-structure can be considered as a large defect structure in which larger air holes are drilled at the central region, the PBG-BE mode can be also compared with the resonant modes of the large defect cavity with five-missing holes along ΓK direction. In our calculation, the quality factors of the modes at such large PC cavities are not large ranging from several hundreds to thousands, which are hundred or ten times smaller than that of the PBG-BE mode. One of the reasons is that the group velocity of the PBG-BE mode, which is close to zero, is much smaller than that of the large cavity modes which experience almost the effective group index of bulk material. If the group velocity of the mode becomes smaller, the scattering loss per unit time would be reduced. Besides, in order to know the origin of the mode, 2D Fourier transform is performed for the PBG-BE mode. The Fourier space field pattern in Fig. 3(d) confirms that the PBG-BE mode originates from the first K point bandedge mode. The wavevectors are localized at the six K points in the Fourier space. Note that the six unparallel wavevectors construct the mode simultaneously. It shows the 2D feedback mechanisms of the first K point bandedge mode.

The threshold of the observed PBG-BE laser was 0.24 mW. We believe that this low threshold is attributed to the high quality factor of the PBG-BE mode. On the other hand, several non-lasing peaks were observed in the spectrum, as shown in Fig. 4(b). The polarizations of all peaks have no definite direction like the first K point bandedge mode. The nonlasing peaks spread over 40 nm near the PBG-BE mode in the frequency domain. Thus, we believe that there may be two possibilities of nonlasing peaks. If any mode satisfies the resonant condition of the five unit-cell cavity and experiences the flat dispersion curve near the bandedge, then the modes can be the resonant mode. Another possibility is that other resonant peaks near the bandedge are created by a little amount of deformation or imperfection in the real sample. The identification of non-lasing peaks is under study.

In conclusion, we have demonstrated the operation of the new type of the bandedge laser, the photonic bandedge laser assisted by the photonic bandgap, in triangular lattice at room temperature. When the air hole radii of the surrounded PC differ from the bandedge mode operating region, most of in-plane loss of the first K point bandedge mode in the central region is suppressed so that the quality factor of the mode become 50000. We identified the PBG-BE lasing modes through the spectral position, CCD image, and the state of polarization, which are matched well with the theoretical predictions. The 2D feedback mechanism of the first K bandedge is proved through the Fourier analysis. The low threshold peak pump power ~ 0.24 mW is achieved due to the low optical loss of the PBG-BE mode.

This work was supported by the National Research Laboratory Project of KISTEP, Korea.

# References

[1] E. Yablonovitch, Phys. Rev. Lett. **58**, 2059 (1987).

[2] S. John, Phys. Rev. Lett. **58**, 2486 (1987).

[3] O. Painter, R. K. Lee, A. Scherer, A. Yariv, J. D. O'Brien, P. D. Dapkus, and I. Kim, Science **284**, 1819 (1999).

[4] S. H. Kim, H. Y. Ryu, H. G. Park, G. H. Kim, Y. S. Choi, Y. H. Lee, and J. S. Kim, Appl. Phys. Lett. **81**, 2499 (2002).

[5] S. H. Kim, G. H. Kim, S. K. Kim, H. G. Park, Y. H. Lee, and S. B. Kim, J. Appl. Phys. **95**, 411 (2004).

[6] J. G. Yang, Y. H. Lee, and S. B. Kim, Appl. Phys. Lett. (to be published 26 April 2004).

[7] H. Altug and J. Vuckovic, Appl. Phys. Lett. **84**, 161 (2004).

[8] S. Noda, M. Yokoyama, M. Imada, A. Chutinan, and M. Mochizuki, Science **293**, 1123 (2001)

[9] H. Y. Ryu, S. H. Kwon, Y. H. Lee, and J. S. Kim, Appl. Phys. Lett. **80**, 3476 (2002).

[10] C. Monat, C. Seassal, X. Letartre, P. Regreny, P. Rojo-Romeo, and P. Viktorovitch, Appl. Phys. Lett. **81**, 5102 (2002).

[11] S. H. Kwon, H. Y. Ryu, G. H. Kim, Y. H. Lee, and S. B. Kim, Appl. Phys. Lett. **83**, 3870 (2003).

[12] J .P. Dowling, M. Scalora, M. J. Bloemer, and C. M. Bowden, J. Appl. Phys. **75**, 1896 (1994)

[13] M. Notomi, H. Suzuki, and T. Tamamura, Appl. Phys. Lett. **78,** 1325 (2001).

[14] H. Y. Ryu, M. Notomi, and Y. H. Lee, Phys. Rev. B **68**, 045209 (2003).

Fig. 1. (a) Top-view scanning electron microscope (SEM) image of a fabricated PBG-BE laser. The lattice constant is 390 nm. The air hole radius of central hexagon region (I) with five holes along the ΓK direction, and it of the surrounded region (II) are 137 nm (0.35$a$) and 117 nm (0.30$a$), respectively. The dotted white hexagon indicates the boundaries of the central region. (b) The band structures calculated by the 2D plane-wave expansion method. The solid lines and the dashed lines correspond to the band structure of air hole radius 0.35$a$ and 0.30$a$. The regions above light line are filled by the dark gray, in which modes are leaky. The transparent gray region indicates the photonic bandgap of the band structure with radius of 0.30$a$. The circle corresponds to the first K point of the band structure with larger air holes, which is the operating point of our PBG-BE laser.

Fig. 2. (Color) (a) Typical mode pattern images of the PBG-BE lasers taken by a CCD camera. The dotted white squares represent the boundary of the PC lasers. (b) The measure polarization states of the laser. (c) The normalized frequency of the PBG-BE laser as a function of air-hole radius (r/$a$) for each lattice constant, a, 325 nm, 345 nm, 365 nm, 390 nm, 414 nm, which are indicated by black, red, green, blue, and cyan, separately. The dots and the solid lines represent the measured lasing positions and the calculated ones by 3D FDTD method.

Fig. 3. (a) Top-view and (b) side-view of the electric field intensity profile of the PBG-BE mode. (c) side-view of the first K point band-edge mode operating at the finite PC pattern without the surrounding air holes. The intensity is indicated by logarithm scale. The white lines show the boundaries of the calculated structures. (d) Fourier space field pattern of the PBG-BE mode.

Fig. 4. (a) Collected output power at the lasing wavelength is plotted for peak pump power. The threshold is ~ 0.24 mW. (b) The spectrum of the PBG-BE lasing mode at peak pump power 0.35mW.

Fig. 1

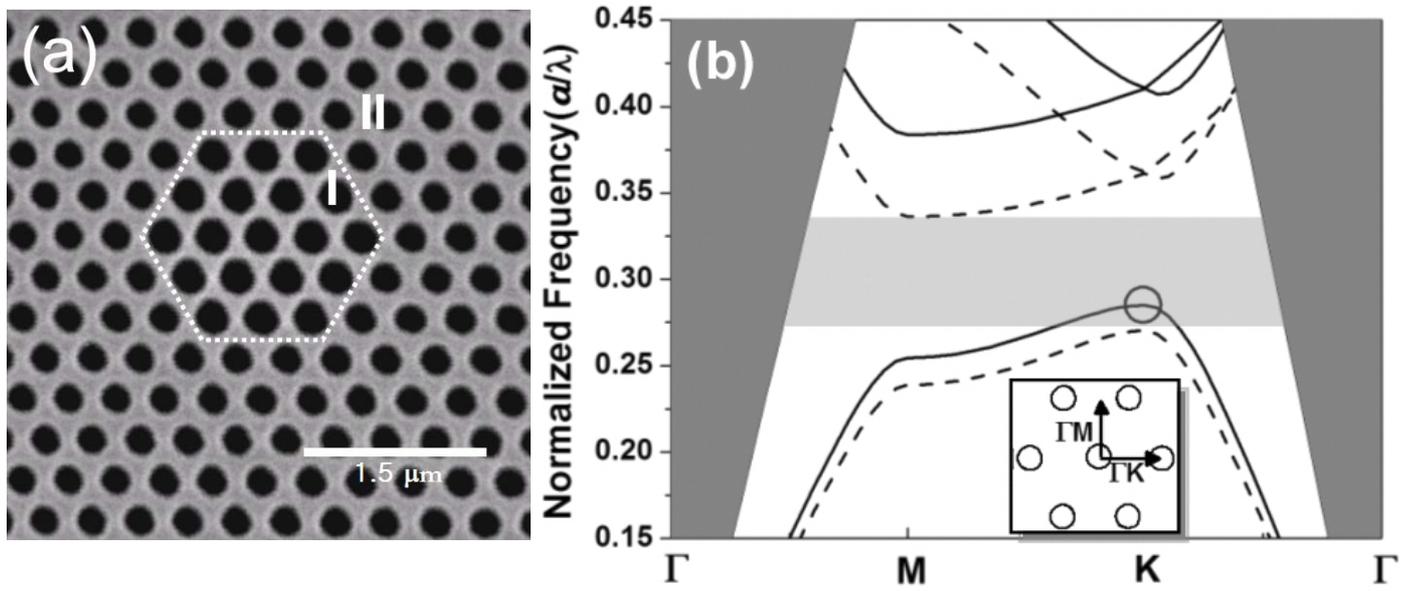

Fig. 2

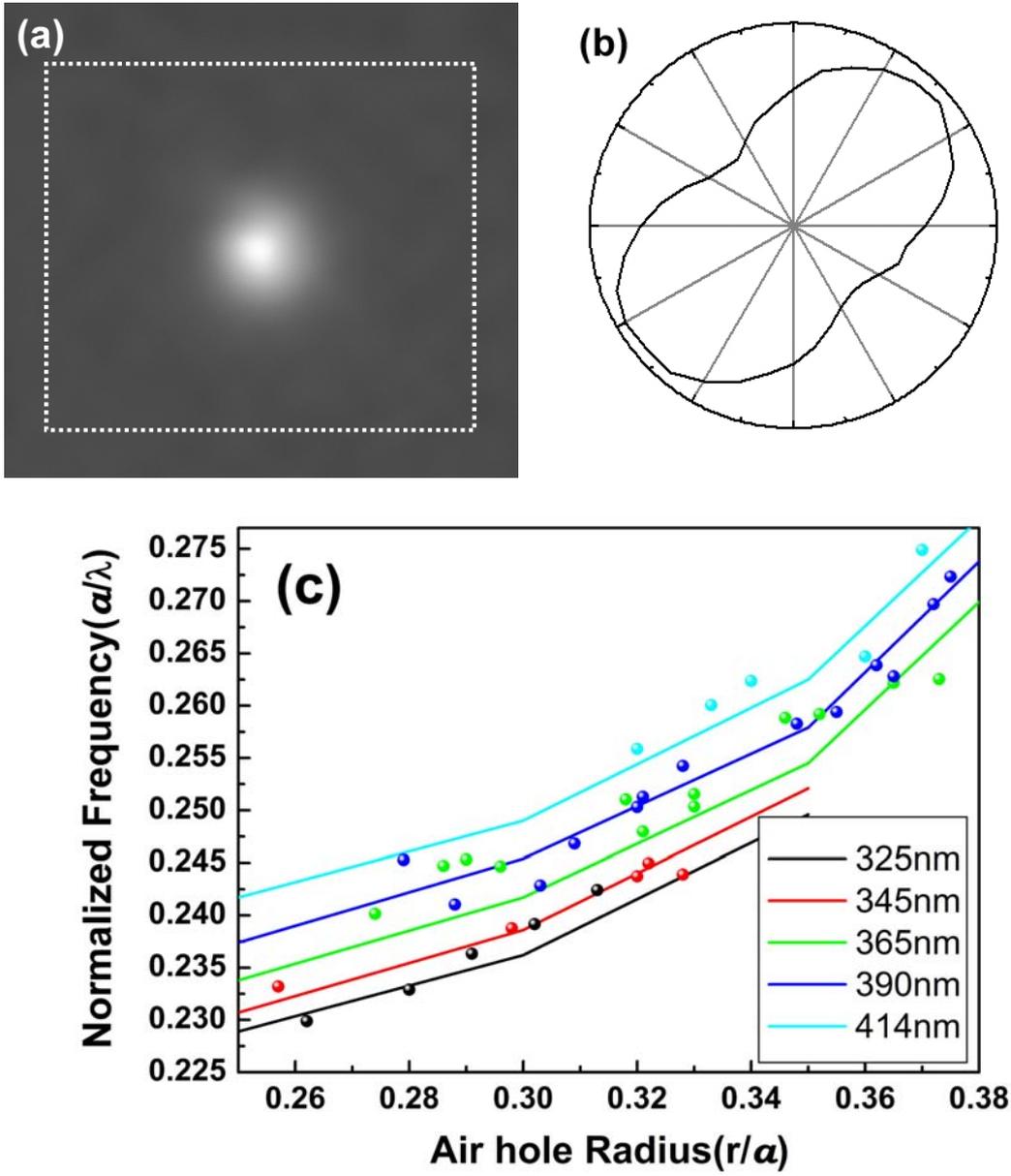

Fig. 3

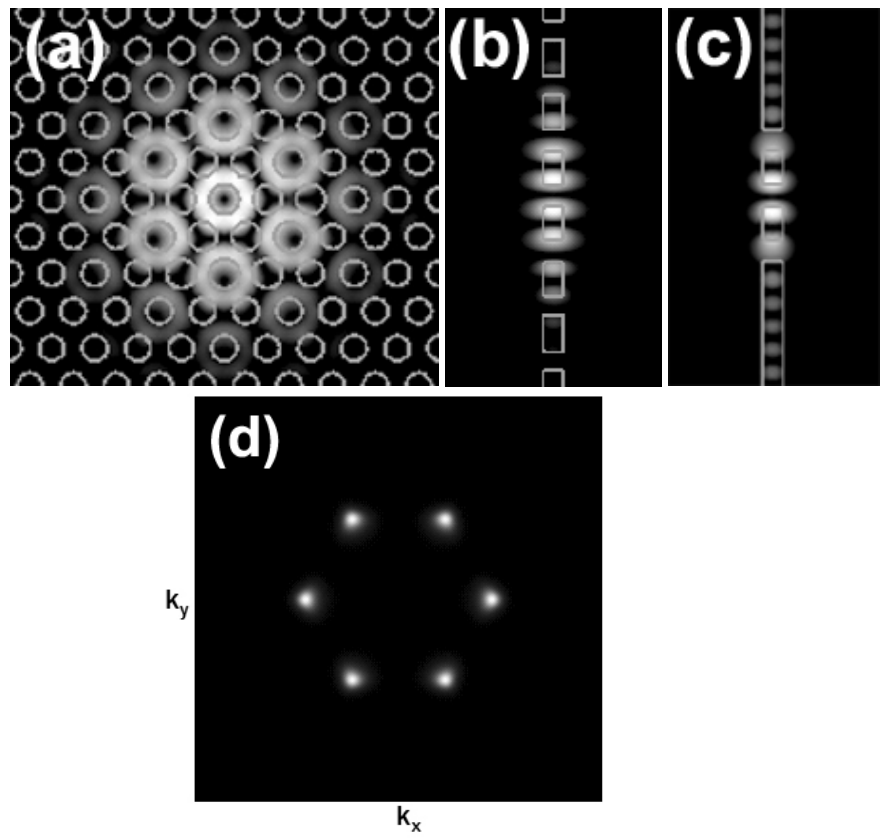

Fig. 4

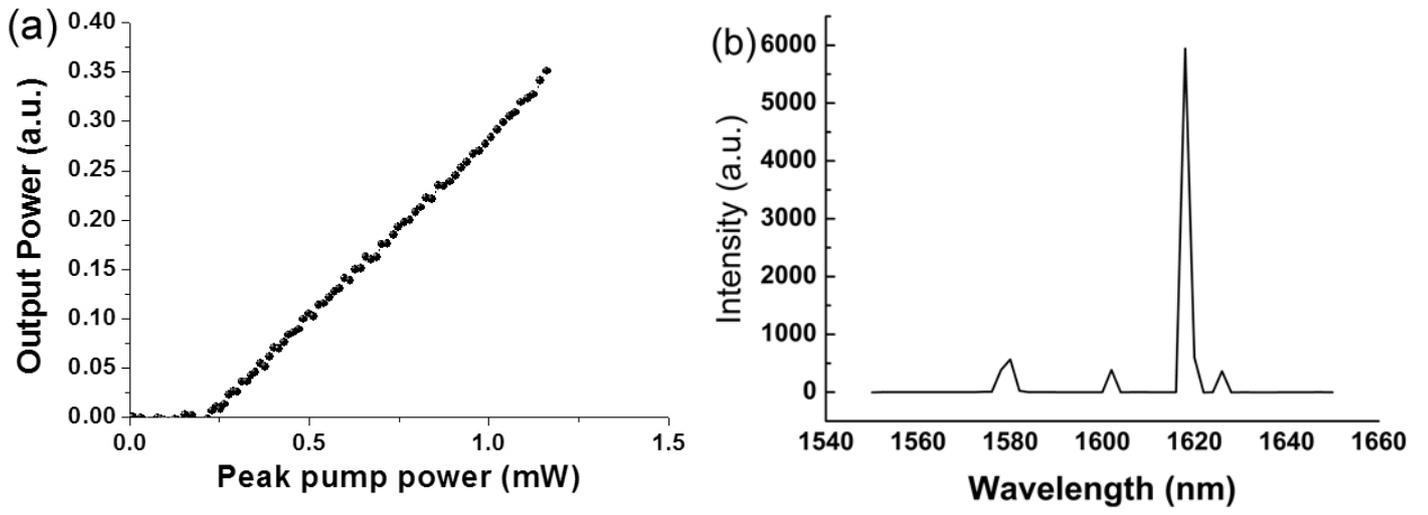